\documentstyle[12pt]{article}

\catcode`\@=11
\long\def\@makefntext#1{ %\parindent 1em
\protect\noindent \hbox to 3.2pt {\hskip-.9pt  
$^{{\ninerm\@thefnmark}}$\hfil}#1\hfill} %can be used 

\def\thefootnote{\fnsymbol{footnote}}
 \def\@makefnmark{\hbox to 0pt{$^{\@thefnmark}$\hss}}  %original 
	
\def\ps@myheadings{\let\@mkboth\@gobbletwo
\def\@oddhead{\hbox{} %\sl
\rightmark\hfil\ninerm\thepage}   
\def\@oddfoot{}\def\@evenhead{\ninerm\thepage\hfil %\sl
\leftmark\hbox{}}\def\@evenfoot{}
\def\sectionmark##1{}\def\subsectionmark##1{}}

\begin{document}

%----------------------------PROCSLA.STY---------------------------------------
\newcommand{\symbolfootnote}{\renewcommand{\thefootnote}
	{\fnsymbol{footnote}}}
\renewcommand{\thefootnote}{\fnsymbol{footnote}}
\newcommand{\alphfootnote}
	{\setcounter{footnote}{0}
	 \renewcommand{\thefootnote}{\sevenrm\alph{footnote}}}

%------------------------------------------------------------------------------
%NEW DEFINED SECTION COMMANDS 
\newcounter{sectionc}\newcounter{subsectionc}\newcounter{subsubsectionc}
\renewcommand{\section}[1] {\vspace{0.6cm}\addtocounter{sectionc}{1} 
\setcounter{subsectionc}{0}\setcounter{subsubsectionc}{0}\noindent 
	{\bf\thesectionc. #1}\par\vspace{0.4cm}}
\renewcommand{\subsection}[1] {\vspace{0.6cm}\addtocounter{subsectionc}{1} 
	\setcounter{subsubsectionc}{0}\noindent 
	{\it\thesectionc.\thesubsectionc. #1}\par\vspace{0.4cm}}
\renewcommand{\subsubsection}[1]
{\vspace{0.6cm}\addtocounter{subsubsectionc}{1}
	\noindent {\rm\thesectionc.\thesubsectionc.\thesubsubsectionc. 
	#1}\par\vspace{0.4cm}}
\newcommand{\nonumsection}[1] {\vspace{0.6cm}\noindent{\bf #1}
	\par\vspace{0.4cm}}
					         
%NEW MACRO TO HANDLE APPENDICES
\newcounter{appendixc}
\newcounter{subappendixc}[appendixc]
\newcounter{subsubappendixc}[subappendixc]
\renewcommand{\thesubappendixc}{\Alph{appendixc}.\arabic{subappendixc}}
\renewcommand{\thesubsubappendixc}
	{\Alph{appendixc}.\arabic{subappendixc}.\arabic{subsubappendixc}}

\renewcommand{\appendix}[1] {\vspace{0.6cm}
        \refstepcounter{appendixc}
        \setcounter{figure}{0}
        \setcounter{table}{0}
        \setcounter{equation}{0}
        \renewcommand{\thefigure}{\Alph{appendixc}.\arabic{figure}}
        \renewcommand{\thetable}{\Alph{appendixc}.\arabic{table}}
        \renewcommand{\theappendixc}{\Alph{appendixc}}
        \renewcommand{\theequation}{\Alph{appendixc}.\arabic{equation}}
%       \noindent{\bf Appendix \theappendixc. #1}\par\vspace{0.4cm}}
        \noindent{\bf Appendix \theappendixc #1}\par\vspace{0.4cm}}
\newcommand{\subappendix}[1] {\vspace{0.6cm}
        \refstepcounter{subappendixc}
        \noindent{\bf Appendix \thesubappendixc. #1}\par\vspace{0.4cm}}
\newcommand{\subsubappendix}[1] {\vspace{0.6cm}
        \refstepcounter{subsubappendixc}
        \noindent{\it Appendix \thesubsubappendixc. #1}
	\par\vspace{0.4cm}}

%------------------------------------------------------------------------------
%MARCO FOR ABSTRACT BLOCK
\def\abstracts#1{{
	\centering{\begin{minipage}{30pc}\tenrm\baselineskip=12pt\noindent
	\centerline{\tenrm ABSTRACT}\vspace{0.3cm}
	\parindent=0pt #1
	\end{minipage} }\par}} 

%------------------------------------------------------------------------------
%NEW MACRO FOR BIBLIOGRAPHY
\newcommand{\bibit}{\it}
\newcommand{\bibbf}{\bf}
\renewenvironment{thebibliography}[1]
	{\begin{list}{\arabic{enumi}.}
	{\usecounter{enumi}\setlength{\parsep}{0pt}
%1.25cm IS STRICTLY FOR PROCSLA.TEX ONLY
\setlength{\leftmargin 1.25cm}{\rightmargin 0pt}
%0.52cm IS FOR NEW DATA FILES
%\setlength{\leftmargin 0.52cm}{\rightmargin 0pt}
	 \setlength{\itemsep}{0pt} \settowidth
	{\labelwidth}{#1.}\sloppy}}{\end{list}}

%------------------------------------------------------------------------------
%FOLLOWING THREE COMMANDS ARE FOR 'LIST' COMMAND.
\topsep=0in\parsep=0in\itemsep=0in
\parindent=1.5pc

%LIST ENVIRONMENTS
\newcounter{itemlistc}
\newcounter{romanlistc}
\newcounter{alphlistc}
\newcounter{arabiclistc}
\newenvironment{itemlist}
    	{\setcounter{itemlistc}{0}
	 \begin{list}{$\bullet$}
	{\usecounter{itemlistc}
	 \setlength{\parsep}{0pt}
	 \setlength{\itemsep}{0pt}}}{\end{list}}

\newenvironment{romanlist}
	{\setcounter{romanlistc}{0}
	 \begin{list}{$($\roman{romanlistc}$)$}
	{\usecounter{romanlistc}
	 \setlength{\parsep}{0pt}
	 \setlength{\itemsep}{0pt}}}{\end{list}}

\newenvironment{alphlist}
	{\setcounter{alphlistc}{0}
	 \begin{list}{$($\alph{alphlistc}$)$}
	{\usecounter{alphlistc}
	 \setlength{\parsep}{0pt}
	 \setlength{\itemsep}{0pt}}}{\end{list}}

\newenvironment{arabiclist}
	{\setcounter{arabiclistc}{0}
	 \begin{list}{\arabic{arabiclistc}}
	{\usecounter{arabiclistc}
	 \setlength{\parsep}{0pt}
	 \setlength{\itemsep}{0pt}}}{\end{list}}

%------------------------------------------------------------------------------
%FIGURE CAPTION
\newcommand{\fcaption}[1]{
        \refstepcounter{figure}
        \setbox\@tempboxa = \hbox{\tenrm Fig.~\thefigure. #1}
        \ifdim \wd\@tempboxa > 6in
           {\begin{center}
        \parbox{6in}{\tenrm\baselineskip=12pt Fig.~\thefigure. #1 }
            \end{center}}
        \else
             {\begin{center}
             {\tenrm Fig.~\thefigure. #1}
              \end{center}}
        \fi}

%TABLE CAPTION
\newcommand{\tcaption}[1]{
        \refstepcounter{table}
        \setbox\@tempboxa = \hbox{\tenrm Table~\thetable. #1}
        \ifdim \wd\@tempboxa > 6in
           {\begin{center}
        \parbox{6in}{\tenrm\baselineskip=12pt Table~\thetable. #1 }
            \end{center}}
        \else
             {\begin{center}
             {\tenrm Table~\thetable. #1}
              \end{center}}
        \fi}

%------------------------------------------------------------------------------
%ACKNOWLEDGEMENT: this portion is from John Hershberger
\def\@citex[#1]#2{\if@filesw\immediate\write\@auxout
	{\string\citation{#2}}\fi
\def\@citea{}\@cite{\@for\@citeb:=#2\do
	{\@citea\def\@citea{,}\@ifundefined
	{b@\@citeb}{{\bf ?}\@warning
	{Citation `\@citeb' on page \thepage \space undefined}}
	{\csname b@\@citeb\endcsname}}}{#1}}

\newif\if@cghi
\def\cite{\@cghitrue\@ifnextchar [{\@tempswatrue
	\@citex}{\@tempswafalse\@citex[]}}
\def\citelow{\@cghifalse\@ifnextchar [{\@tempswatrue
	\@citex}{\@tempswafalse\@citex[]}}
\def\@cite#1#2{{$\null^{#1}$\if@tempswa\typeout
	{IJCGA warning: optional citation argument 
	ignored: `#2'} \fi}}
\newcommand{\citeup}{\cite}

%------------------------------------------------------------------------------
%FOR FNSYMBOL FOOTNOTE AND ALPH{FOOTNOTE} 
\def\fnm#1{$^{\mbox{\scriptsize #1}}$}
\def\fnt#1#2{\footnotetext{\kern-.3em
	{$^{\mbox{\sevenrm #1}}$}{#2}}}

%------------------------------------------------------------------------------
\font\twelvebf=cmbx10 scaled\magstep 1
\font\twelverm=cmr10 scaled\magstep 1
\font\twelveit=cmti10 scaled\magstep 1
\font\elevenbfit=cmbxti10 scaled\magstephalf
\font\elevenbf=cmbx10 scaled\magstephalf
\font\elevenrm=cmr10 scaled\magstephalf
\font\elevenit=cmti10 scaled\magstephalf
\font\bfit=cmbxti10
\font\tenbf=cmbx10
\font\tenrm=cmr10
\font\tenit=cmti10
\font\ninebf=cmbx9
\font\ninerm=cmr9
\font\nineit=cmti9
\font\eightbf=cmbx8
\font\eightrm=cmr8
\font\eightit=cmti8

%----------my macros and other inputs
\input epsf
\def\epsfsize#1#2{0.60#1}

\def\OMIT#1{}
\def\np#1#2#3{Nucl. Phys. {\bf #1} (#2) #3}
\def\pl#1#2#3{Phys. Lett. {\bf #1} (#2) #3}
\def\prl#1#2#3{Phys. Rev. Lett. {\bf #1} (#2) #3}
\def\pr#1#2#3{Phys. Rev. {\bf #1} (#2) #3}
\def\sjnp#1#2#3{Sov. J. Nucl. Phys. {\bf #1} (#2) #3}
\def\blankref#1#2#3{   {\bf #1} (#2) #3}
\def\vsl{v \hskip-5pt /}
\def\ie{{\it i.e.}}
\def\eg{{\it e.g.}}
\def\cf{{\it cf., }}
\def\ccdot{\hbox{\kern-.1em$\cdot$\kern-.1em}}
\def\vv{v \ccdot v'}
\def\LB{\Lambda_b}
\def\LC{\Lambda_c}
\def\bas{\bar\alpha_s}
\def\bv{b_v}
\def\cvp{\overline{c}_{v'}}
\def\Dl{\overleftarrow{D}}
\def\Dlslash{{\overleftarrow{D}} \hskip-0.75 em / \hskip+0.40 em}
\def\Dslash{D\hskip-0.65 em / \hskip+0.30 em}
\def\gf{\gamma^5}
\def\gu{\gamma^\mu}
\def\imb{\displaystyle{i \over m_b}}
\def\imc{\displaystyle{i \over m_c}}
\def\imu{i \mu^\epsilon\,}
\def\mb{m_b}
\def\mc{m_c}
\def\muep{\mu^\epsilon}
\def\muhalf{\mu^{\epsilon /2}}
\def\suv{\sigma^{\mu\nu}}
 \def\sl#1{#1\hskip-0.5em /}  % Mike Luke's macro for slashes through vectors.
\def\v{v^\mu}
\def\vp{v'^\mu}
\def\Asl{\hbox{/\kern-.6500em \rm A}}
\def\Dsl{\hbox{/\kern-.6000em\rm D}} %Roman D slash
\def\dsl{\,\raise.15ex\hbox{/}\mkern-13.5mu D} %D slash
%
% Miscellaneous macros

\def\A{{\cal A}}
\def\B{{\cal B}}
\def\M{{\cal M}}
\def\gpiBB{g_{\pi B n}}
\def\gpiBBo{g_{\pi B B}}
\def\gn{{f_{n}}}

\def\tL{\tilde\Lambda}
\def\d{{\rm d}}
\def\qm{q^2_{\rm max}}
\def\o{\omega}
\def\voo{\left({v\over\o}\right)}
\def\voom{\left({v\over\o_-}\right)}
\def\vmo{\left({v-\o\over1-\o}\right)}

\def\PI{\pi}
\def\refmark#1{[#1]}
\def\art#1{{\sl ``#1''}}

\def\pvint{-\mskip-19mu\int}
\def\d{{\rm d}}
\def\dv{{\rm d}v\,}
\def\dx{{\rm d}x\,}
\def\dy{{\rm d}y\,}

\def\PHI#1#2{\phi_{#1}( {#2} )}
\def\GAMMA#1#2{\Phi_{#1}( {#2} )}
\def\OMIT#1{}

\def\vev#1{{\left\langle #1 \right\rangle}}
\def\D{{\rm d}}

\def\ie{{\it i.e.}}
\def\eg{{\it e.g.}}

\def\AP{{\it Ann.\ Phys.\ }}
\def\CMP{{\it Comm.\ Math.\ Phys.\ }}
\def\CTP{{\it Comm.\ Theor.\ Phys.\ }}
\def\IJMP{{\it Int.\ Jour.\ Mod.\ Phys.\ }}
\def\JETPL{{\it JETP Lett.\ }}
\def\NC{{\it Nuovo Cimento\ }}
\def\NP{{\it Nucl.\ Phys.\ }}
\def\PL{{\it Phys.\ Lett.\ }}
\def\PR{{\it Phys.\ Rev.\ }}
\def\PRep{{\it Phys.\ Rep.\ }}
\def\PRL{{\it Phys.\ Rev.\ Lett.\ }}
\def\SJNP{{\it Sov.~J.~Nucl.~Phys.\ }}

\def\ctp#1#2#3{\CTP{\bf #1} (#2) #3}
\def\nc#1#2#3{\NC{\bf #1} (#2) #3}
\def\np#1#2#3{\NP{\bf #1} (#2) #3}
\def\pl#1#2#3{\PL{\bf #1} (#2) #3}
\def\prl#1#2#3{\PRL{\bf #1} (#2) #3}
\def\pr#1#2#3{\PR{\bf #1} (#2) #3}
\def\sjnp#1#2#3{\SJNP{\bf #1} (#2) #3}
\def\nuvc#1#2#3{\NC{\bf #1A} (#2) #3}
\def\blankref#1#2#3{   {\bf #1} (#2) #3}
\def\ibid#1#2#3{{\it ibid,\/}  {\bf #1} (#2) #3}

\def\Exp#1{{\rm e}^{#1}}
\def\eqsim{\cong}

\newcommand{\eqn}[2]{\begin{equation} #2 \label{#1}\end{equation}}
\newcommand{\eqna}[1]{\begin{eqnarray} #1 \end{eqnarray}}
%\newcommand{\vev}[1]{\langle #1 \rangle}
%\def\vev#1{\langle #1 \rangle}
%----------------------START OF DATA FILE------------------------------
\baselineskip=22pt
\begin{titlepage}
\begin{flushright}
SMU-HEP/94-11\\
UCSD/PTH 94-09
\end{flushright}
\vspace{1cm}
\begin{center}{
\LARGE Exact Heavy To Light Meson Form Factors In 
The Chiral Limit Of
Planar 1+1  QCD\footnote{Talk Presented at the Workshop on Continuous
Advances in QCD, Theoretical Physics Institute, University of Minnesota,
February 18--20, 1994}}\\
\vspace{1cm}
{\large Benjam\'\i n Grinstein\footnote{e-mail:ben@mail.physics.smu.edu}}\\
\baselineskip=14pt
\vspace{1cm}
Department of Physics, Southern Methodist University\\
 Dallas, Texas 75275, USA\\
and\\
Department of Physics, University of California,
San Diego\footnote{Address after July 1, 1994}, \\
La Jolla, California, 92093-0319, USA\\
\parbox{5.5in}{\vspace{1in}\begin{center}ABSTRACT\end{center}}
\parbox{5.5in}{ 
The form factors of the flavor changing vector current between a
$\bar q Q$-meson and the lightest $\bar q q$ (pseudoscalar-)meson are
computed exactly and explicitly in the 't~Hooft model (planar QCD in
$1+1$ dimensions) in the limit that the mass of $q$-quark vanishes.
\vspace{1in}}
\end{center}
May 20, 1994
\end{titlepage}

\centerline{\tenbf EXACT HEAVY TO LIGHT MESON FORM FACTORS }
\baselineskip=22pt
\centerline{\tenbf IN THE CHIRAL LIMIT OF PLANAR 1+1  QCD.}
\baselineskip=16pt
\vspace{0.8cm}
\centerline{\tenrm BENJAMIN GRINSTEIN}
\baselineskip=13pt
\centerline{\tenit Department of Physics, Southern Methodist University,
 Dallas, Texas 75275, USA}
%\vspace{0.3cm}
\centerline{\tenrm and}
%\vspace{0.3cm}
\centerline{\tenit Department of Physics, University of California at 
San Diego, La Jolla, California, 92093-0319, USA}
\vspace{0.9cm}
\abstracts{The form factors of the flavor changing vector current
between a $\bar q Q$-meson
and the  lightest $\bar q q$ (pseudoscalar-)meson  are computed exactly and
explicitly in the 't~Hooft model (planar QCD in $1+1$ dimensions)
in the limit that the mass of $q$-quark vanishes.
}

\vfil
%\vspace{0.8cm}
\rm\baselineskip=14pt
\section{Introduction}

Little is known about form factors of local operators between
a heavy meson like the~$\bar B$ --- with quantum numbers of
a single heavy quark~$Q$  and a
single light antiquark~$\bar q$ --- and light pseudoscalar mesons like the
$\pi$--$K$--$\eta$ octet.
Isgur and Wise have shown that heavy quark symmetries\cite{IWb}
relate several form factors\cite{IWa},
but nothing is known about their shape.
Thus far all theoretical attempts to describe them are based on
particular models of hadrons.

Surprisingly one can calculate the
shape of these form factors exactly in the chiral limit of planar
QCD in 1+1 dimensions. As we will see, the form factors $f_\pm$ (defined
below) are
given by a single pole at the mass of the $B$ with residue 
$\pm \mu^2_B f_B/f_\pi$, where $\mu_X$ and $f_X$ are 
 the mass and decay constant of the $X$-meson, respectively.
Whether this result is indicative of the behavior of the form
factors in 3+1 QCD is at present a matter of pure speculation.

This talk reports on work done in collaboration with Paul Mende, first
reported elsewhere.\cite{mendebg} The shape of the form factor will be
determined in section~2, and its normalization in section~3. These
analytic results are compared with numerical computations in section~4.

\section{Proof of Pole Dominance}

The matrix element of the vector current  $V_\nu=\bar q\gamma_\nu Q$ can be
expressed in terms of two form factors~$f_\pm$: 
\eqn{ffsdefd}{
\vev{\PI(p') | V_\nu | B(p)} = (p+p')_\nu f_+(q^2) + (p-p')_\nu f_-
(q^2)~.
}
where $q=p-p'$ and throughout the paper $p$ and~$p'$
alway denote the momentum of the~$B$ and~$\pi$ mesons,
respectively.
In 1+1 dimensions by  ``$B$ and
$\PI$ mesons'' we mean that they are the lightest states with quantum numbers 
$Q\bar q$ and $q\bar q$, respectively. In what follows we
will take the quark~$Q$ to have a fixed mass~$M$ and the quark~$q$ to have
mass~$m$, and we will consider the limit $m\to 0$. 

In planar (ie, large $N_c$) QCD the form factors in (\ref{ffsdefd})
 are saturated by
couplings of the flavor changing current~$V_\nu$ to the $Q\bar q$ resonances
in that channel.\cite{thooft,CCG,einhorn} We therefore can write
   \eqn{Forms}{
   \vev{\PI | V_\nu | B}
   = \sum_n {\vev{0|V_\nu|B_n}  \vev{\PI B_n|B}\over q^2 - \mu_n^2 }
   .}
Here $B_n$ denote the states with quantum number $\bar q Q$ ordered by
increasing mass $\mu_n$.

For odd parity states $|B_n\rangle$,
   \eqn{oddfdef}{
   \vev{0|V_\nu|B_n} = \epsilon_{\nu\lambda}q^\lambda \gn
   .}
We can describe the interactions giving the matrix element $\vev{\PI
B_n|B}$ conveniently in terms of
a hadron lagrangian, dual to the fundamental lagrangian,
which couples the mesons via terms
   \eqn{Lodd}{
   {\cal L}_{\rm int}=
   \sum_{abc}
   \hat g_{abc}(q^2)\epsilon_{\lambda\nu}
   \, \partial_\lambda \phi^a \partial_\nu \phi^b \phi^c~.
   }
Similarly,  for even parity,
\eqn{evenfdef}{
   \vev{0|V_\nu|B_n} = q_\nu \gn
   ,}
with couplings
   \eqn{Leven}{
   {\cal L}_{\rm int}=
   \sum_{abc}
    \hat g_{abc}(q^2) \, \phi^a\phi^b\phi^c.
   }

The form factors of Eq.~(\ref{ffsdefd}) can then be written
\eqna{
%   \eqalign{
   f_+ &=& \sum_{{\rm even\ parity}}
   {- \gn(q^2) \hat\gpiBB(q^2) q^2 \over q^2-\mu_n^2}
   \\
   f_- &=& \sum_{{\rm odd\ parity}} 
         { \gn(q^2) \hat\gpiBB(q^2) \over q^2-\mu_n^2}
   + \sum_{{\rm even\ parity}}
   { \gn(q^2) \hat\gpiBB(q^2) (\mu_B^2-\mu_\pi^2) \over q^2-\mu_n^2}
   ~, \label{ffexpn}
%}
}
which is obtained with the help of the useful
two-dimensional formula
\eqn{useful}{
   \epsilon_{\lambda\nu}q^\nu
   = \left\lbrack -q^2(p+p')_\lambda + (\mu_B^2 - \mu_\pi^2)(p-p')_\nu
   \right\rbrack / 2\epsilon^{\rho\sigma}p_\rho p'_\sigma ~.
}
Note that the expansions~(\ref{ffexpn}) have momentum dependent
numerators, proportional to the off-shell decay constants $\gn(q^2)$
and three point
couplings, $\hat\gpiBB(q^2)$. If the form factors vanish as $|q^2|\to\infty$, 
we can replace the momentum dependent numerators by on-shell numerators,  
\eqna{
%   \eqalign{
   f_+ &=& \sum_{{\rm even\ parity}}
   {- \gn \hat\gpiBB(\mu_n^2) \mu_n^2 \over q^2-\mu_n^2}
   \\
   f_- &=& \sum_{{\rm odd\ parity}}
   { \gn \hat\gpiBB(\mu_n^2) \over q^2-\mu_n^2}
   + \sum_{{\rm even\ parity}}
   { \gn \hat\gpiBB(\mu_n^2) (\mu_B^2-\mu_\pi^2) \over q^2-\mu_n^2}
   ~, \label{ffexpntoo}
%}
}
\begin{figure}
\hskip1.3in\epsfbox{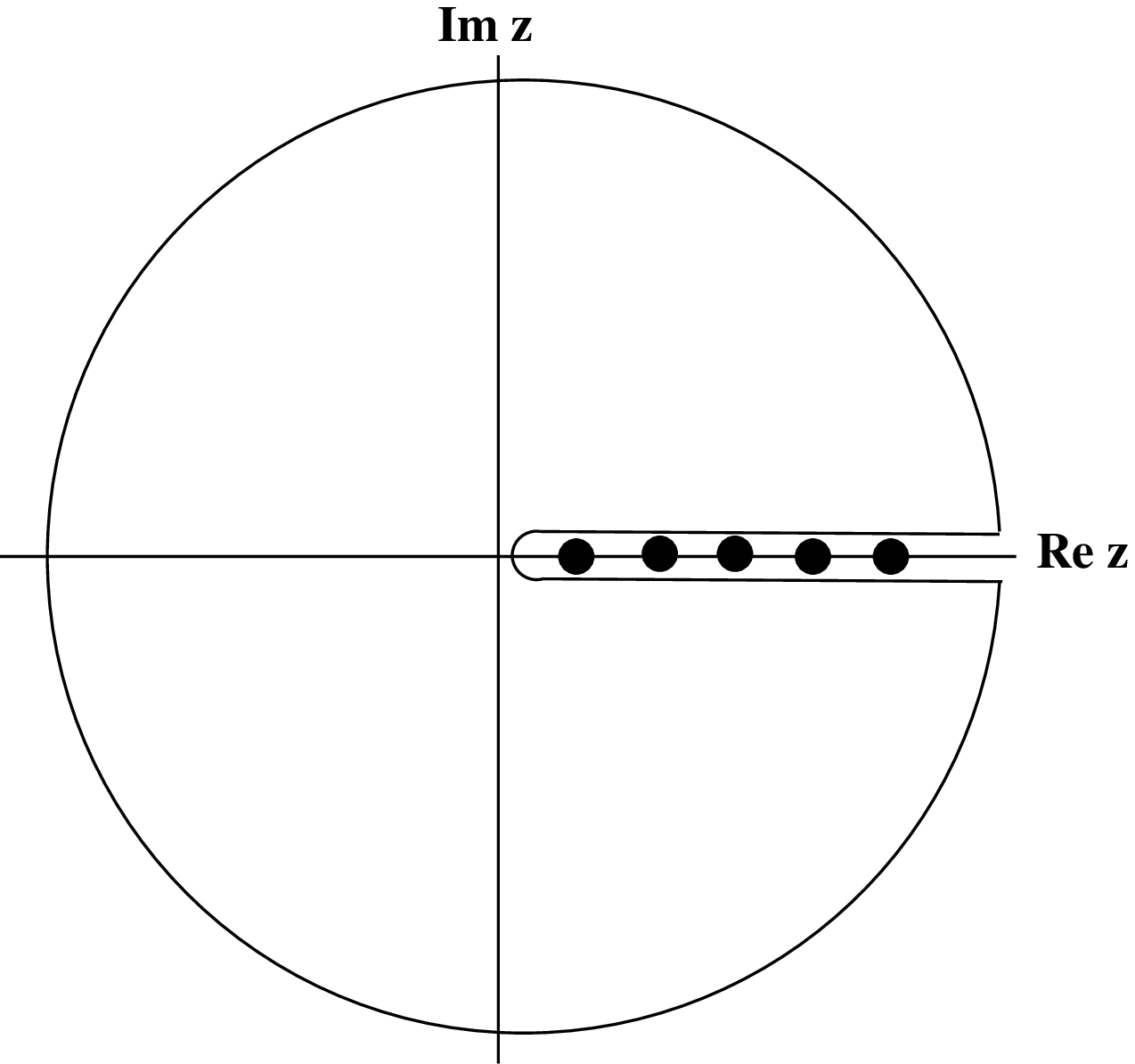}
\fcaption{Contour of integration for the integral in Eq.~\ref{cauchy}.
The filled circles denote the location of poles of the form factors, 
$z=\mu_0^2,\mu_1^2, \mu_2^2,\ldots$}
\label{fig:contour}
\end{figure}

To verify this, consider Cauchy's theorem for the form factors, 
\eqn{cauchy}{
f_\pm(q^2) = {1\over2\pi i}\int_C dz {f_\pm(z)\over z-q^2}
}
where the contour $C$ consists of a circle $|z|=R\to\infty$, excluding the
positive real axis, $z=R$, and a line just below and above the
positive real axis extending from $z = \mu_B^2 \pm i\epsilon$ to $z =
R \pm i\epsilon\to\infty \pm i\epsilon$ (and enclosing $\mu_B^2$ on
the left); see Fig.~\ref{fig:contour}. 
Now, if, as assumed, $|f_\pm(z)|\to0$ as $|z|\to\infty$,
the integral over the circle at infinity vanishes. The integral can
thus be traded for a sum over integrals over infinitesimal contours
around each pole, $z=\mu_n^2$.  Using the explicit form of $f_\pm$ in
Eq.~(\ref{ffexpn}), we see that only the $n$-th term in the sum has a pole at
$z=\mu_n^2$. Applying Cauchy's theorem again gives the on-shell
numerators.

Several observations are in order.
It is large $N_c$ which allows us to treat the resonances as stable
without continuum couplings in Eq.~(\ref{Forms}).
It selects the three point couplings
to one-particle intermediate states.
In using convergence as $|q^2|\to \infty$,
we make an assumption about the large-momentum behavior
of the interactions, information unavailable from
a low-energy analysis or standard chiral lagrangian analysis.
The shift of the numerators to the residues at the poles,
familiar in dispersion theory, has a simple physical origin:
in the ``effective'' meson field theory there is
freedom to make arbitrary field redefinitions without
changing the on-shell $S$-matrix.
Here that freedom is used to replace the momentum dependent couplings
(that is, the higher derivative operators)
by constants at the expense of
shifting the coefficients of higher point functions, which in
turn are down by powers of $1/N_c$.
Note the contrast with the use of field redefinitions by 
Georgi\citelow{georgi}. In his analysis higher point functions are
suppressed instead by powers of the cutoff. The 
required behaviour as $|q^2|\to\infty$ can be established by explicit analysis
of the quark scattering matrix\cite{einhorn}.

Now in the chiral limit --- the light quark mass $m\to 0$
and $\mu_\pi^2\to 0$ ---
the on-shell three-point couplings $\vev{\PI B_n|B}$ vanish
while the decay constants $\gn$ remain finite.
This is a direct consequence of chiral symmetry and leads immediately
to a pole-dominated form factor, ie, $f(q^2)\sim 1/(q^2-\mu_B^2)$.

To show this in detail, fix the state $B_n$ and consider
the matrix element of the {\it light-light\/} current
$a_\lambda=\bar q\gamma_\lambda \gamma_5 q$:
   \eqn{Lightlight}{
   \vev{B_n| a_\lambda |B}
   = \sum_\ell { \vev{ 0 | a_\lambda | \PI_\ell} \vev{\PI_\ell B_n|B}
   \over p'^2 - \mu_\ell^2 }
   }
Here the sum is over the tower of $\bar q q$ states,
$|\PI_\ell(p')\rangle$. Again, since the form factors of $a_\lambda$
vanish\cite{einhorn} as $|p'^2|\to\infty$, we replace the numerators
in the sum in~(\ref{Lightlight}) by their on-shell values.

Observe that axial current conservation implies
$0=\vev{ 0 | \partial\cdot a | \PI_\ell} =\mu_\ell^2f_\ell$. Therefore
$f_\ell=0$ unless $\mu_\ell=0$, so  the axial current couples only to the
massless pion,  $\PI=\PI_0$. Therefore in the chiral limit,
\eqn{Lightlimit}{
  \vev{B_n| a_\lambda |B}
					\to { f_\pi \vev{\pi B_n|B}
   \over p'^2 }p'_\lambda
	}

Again applying axial current conservation, we have
   \eqn{Axial}{
   0=\partial\cdot a \to {f_\pi \vev{\pi B_n|B}\over p'^2}p'^2 
 }
or, more explicitly,
\eqn{explicitlight}{
 0
   = \cases{
   f_\pi \hat\gpiBB(\mu_n^2) & if $n$ is odd; \cr
   (2\epsilon^{\lambda\sigma}p_\lambda p'_\sigma)f_\pi \hat\gpiBB(\mu_n^2)
    & if $n$ is even.
   \cr}
   }
We emphasize that this equation is valid when all three states are on-shell.
 It immediately follows that
   \eqn{Vanish}{
   \hat\gpiBB (\mu_n^2)= 0,
   \qquad n \ne 0
   .}
so, except for the ground state, all the coupling constants in the
effective lagrangian in~(\ref{Lodd}) and~(\ref{Leven}) vanish on the
mass-shell.   The case
$n=0$ is singled out because the factor
$2\epsilon^{\lambda\sigma}p_\lambda p'_\sigma=0$, so it need not
follow that $\hat g_{\pi B B}$ vanishes.  Below we show that it does not.

Combining these results and
introducing the coupling $\gpiBBo=\mu^2_B\hat\gpiBBo(\mu_B^2)$ in
analogy with the definition that is natural in four dimensions,
the form factors are
\eqn{twodimffs}{
   f_+= - f_-  
   = -  {f_B\,\gpiBBo\over q^2-\mu_B^2}
  }
%

%%%%%%%%%%%%%%%%%%%%%%%%%%%%%%%%%%
\vfill\eject
\section{Normalization}

The coupling $\gpiBBo$ can be fixed by an application of the Callan-Trieman
relation, adapted to 1+1 QCD. To derive it, consider the
following matrix element,
\eqn{CTstart}{
\M_{\mu\nu} = i\int d^2\! x \, e^{ip'x}\vev{0|T(a_\mu(x) V_\nu(0))|B(p)}
}
In the chiral limit, the light-light axial current $a_\mu$ is conserved. The
divergence $p'_\mu\M_{\mu\nu}$ gives therefore an equal time commutator of
$a_0$ and $V_\nu$. One has
 \eqn{CTdivone}{ 
p'_\mu\M_{\mu\nu} \to -i\vev{0|A_\nu|B(p)}=f_B p_\nu }
Alternatively, evaluate the matrix element and then contract with $p'$. 
The currents $a_\mu$ and $V_\nu$ create states in the $\pi$ and $B$
towers, respectively, and they couple to the $B$ meson through the
matrix element $\vev{\pi_\ell B_n|B}$, thus,
\eqn{doublepoles}{
 \sum_\ell\sum_n {f_{\pi_\ell}p'_\mu\over  p'^2-\mu_\ell^2}
{ f_{n\nu} \over q^2-\mu_n^2}\vev{\pi_\ell B_n|B}
}
where by $f_{n\nu}$ we mean either of~(\ref{oddfdef}) or~(\ref{evenfdef})
according to whether $B_n$ has odd or even parity, respectively. Now,
contracting with $p'_\mu$ and letting $p'\to0$ only the $\ell=0$ term,
that is the pion, remains in the sum over $\ell$. Moreover, all the
matrix elements  $\vev{\pi_\ell B_n|B}$ must vanish linearly with
$p'$, beacuse goldstone bosons are always derivatively coupled. The
only term that survives is the one with $n=0$, that is, the $B$-meson
term itself, because the denominator $q^2-\mu_B^2$ also
vanishes linearly with $p'$. Explicitly we have then
\eqn{preCRT}{
f_Bp_\nu = \lim_{p'\to0}{[f_B(q^2)\epsilon_{\nu\lambda}q^\lambda]
[2\epsilon^{\rho\sigma}p_\rho p'_\sigma \hat\gpiBBo(q^2)]\over
q^2-\mu_B^2}
}
This can be simplified with the use of Eq.~(\ref{useful}). We note  that in the
limit $p'\to0$ one has $q^2=\mu_B^2$ and the couplings involved are
automatically 
on-shell, so we obtain
\eqn{CRT}{
-f_\pi \hat\gpiBBo(\mu_B^2) =1
}
Since $f_\pi=\sqrt{N_c/\pi}$, this completely determines the
coupling~$\gpiBBo$. 

Our final results reads as follows
\eqn{finally}{
f_+=-f_-= { f_B/f_\pi\over q^2/\mu_B^2-1}
}
These are the {\it exact} form fators in the chiral limit.
\vfill \eject

\section{Experimental Verification}
In this section we will assume a unit system with $g^2N_c/\pi=1$.

The results of the previous sections have been verified through
numerical investigations of the case of small, nonvanishing `light
mass' $m$.\cite{mendebg} The form factors may be written as a sum over
resonances,
\eqn{polesinA}{
 f_+(q^2)
   = \sum_{{\rm even}\ n}{\A_n(q^2)\over1- q^2/\mu_n^2}
}
with residues given in terms of the couplings introduced above by
\eqn{Adefd}{
 \A_n(q^2) \equiv - { \gn\gpiBB(q^2) \over \mu_n^2 } ~.
}
\begin{figure}
\hskip0.9in\epsfbox{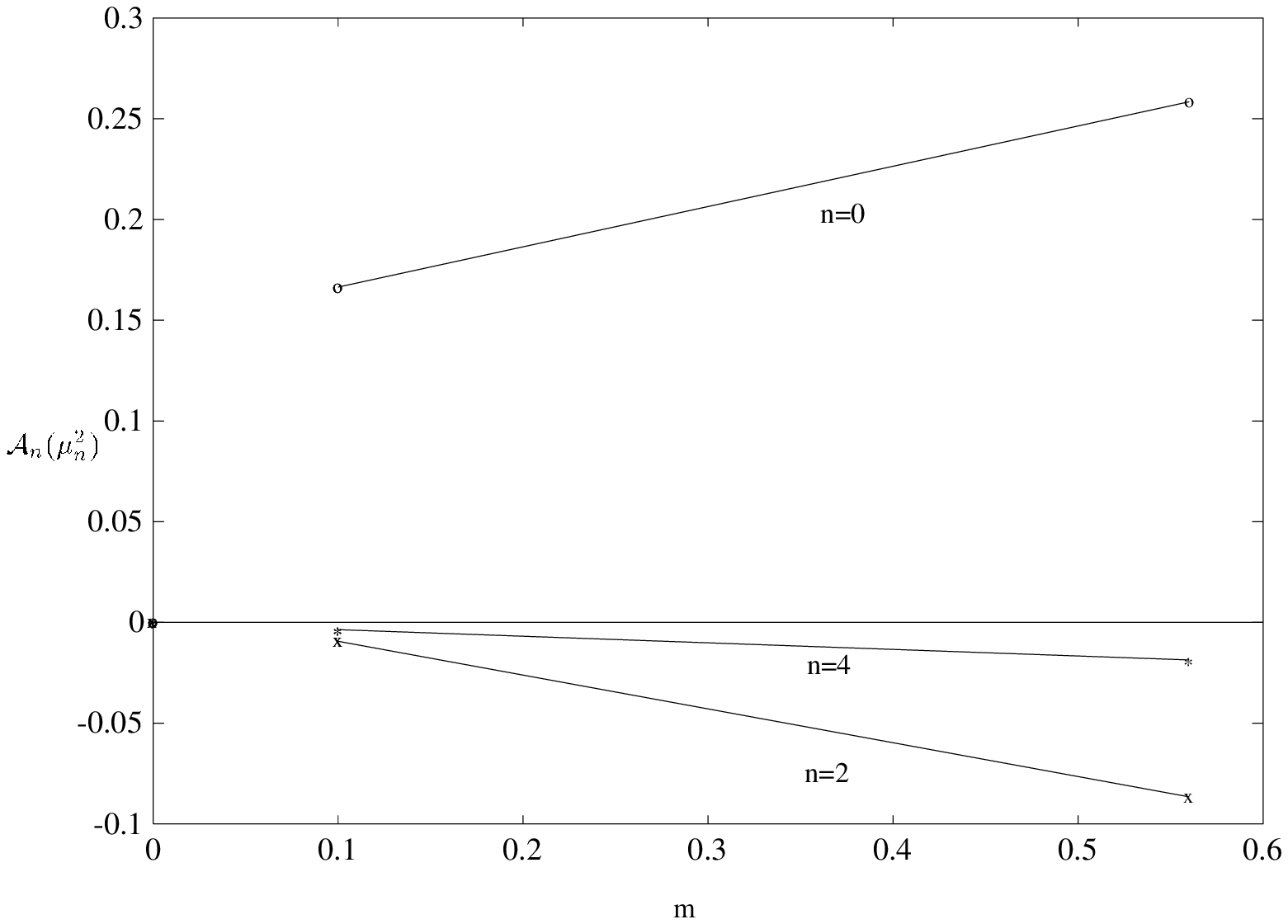}
\fcaption{
The residue $\A_n(\mu_n^2)$ vs. $m$, the light quark mass.
The value was computed numerically for $m=0.1, 0.56$
and a line connecting the pairs of points drawn to guide the eye.
For $n \ne 0$, the $\A_n \to 0$ in the chiral limit, $m\to 0$.
Here $M^2=20000$.
}\label{fig:Avsmone}
\end{figure}

\begin{figure}
\hskip0.9in\epsfbox{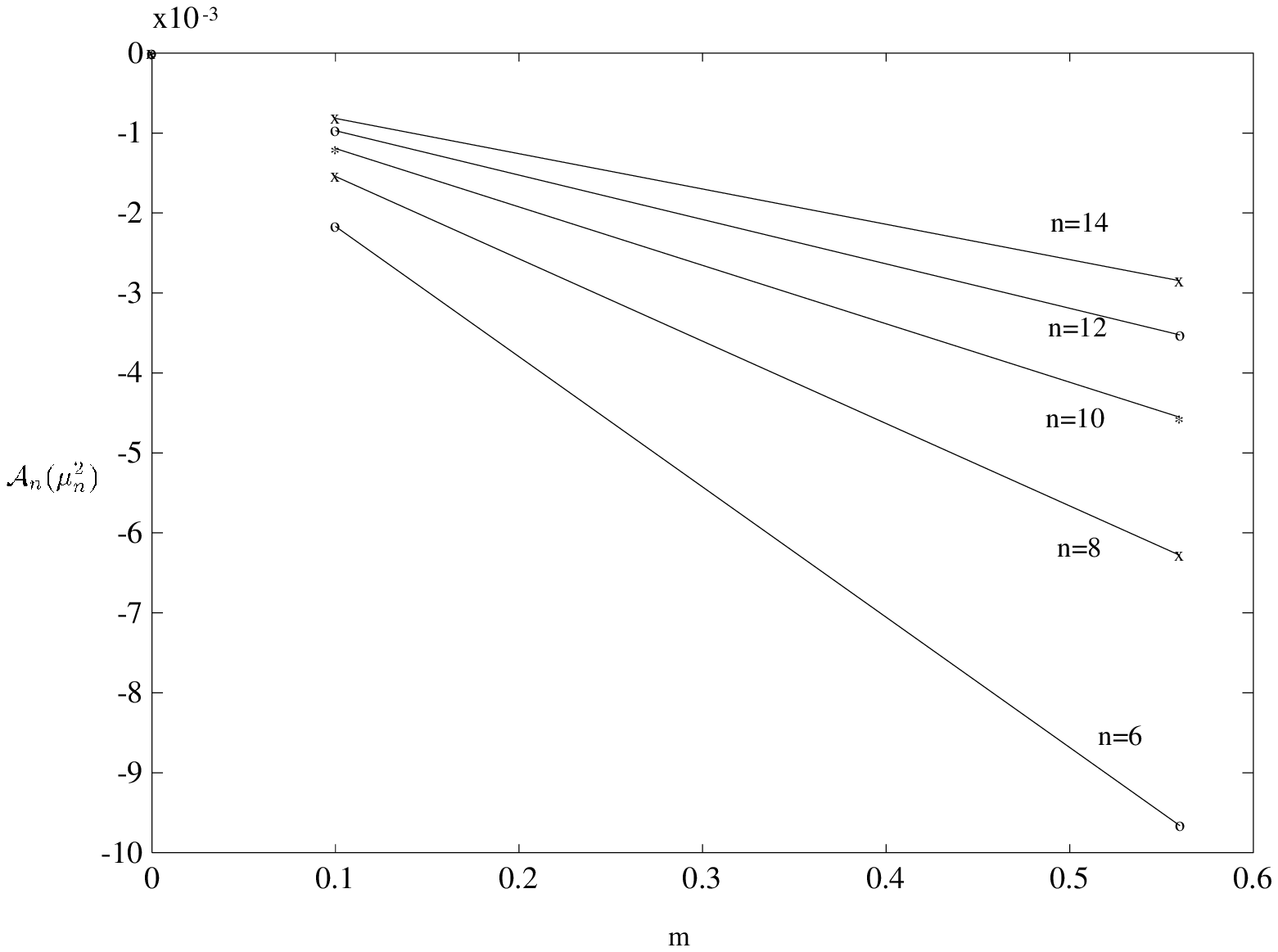}
\fcaption{
The residue $\A_n(\mu_n^2)$ vs. $m$, the light quark mass,
for more resonant states, $n=$6--18.
}
\label{fig:Avsmtwo}
\end{figure}

Figs.~\ref{fig:Avsmone} and~\ref{fig:Avsmtwo} show the residues
$\A_n(\mu_n^2)$ plotted as functions of the light quark mass for $n =
0$ through 14. To perform this computation one first writes the
residues $\A_n(\mu_n^2)$ as overlap integrals of\ 't~Hooft wave
functions\cite{mendebg}. The 't~Hooft wave functions can be determined
numerically, and the overlap integrals can also be done
numerically. It is apparent that $\A_n(\mu_n^2)\to0$ as $m\to0$,
except for $n=0$, which approaches a limit consistent with
Eq.~(\ref{CRT}). The decay constants are rather insensitive to
$m$\cite{burswan}, so the couplings $\hat\gpiBB$ are seen to vanish linearly
with it.

\begin{figure}
\hskip0.7in\epsfbox{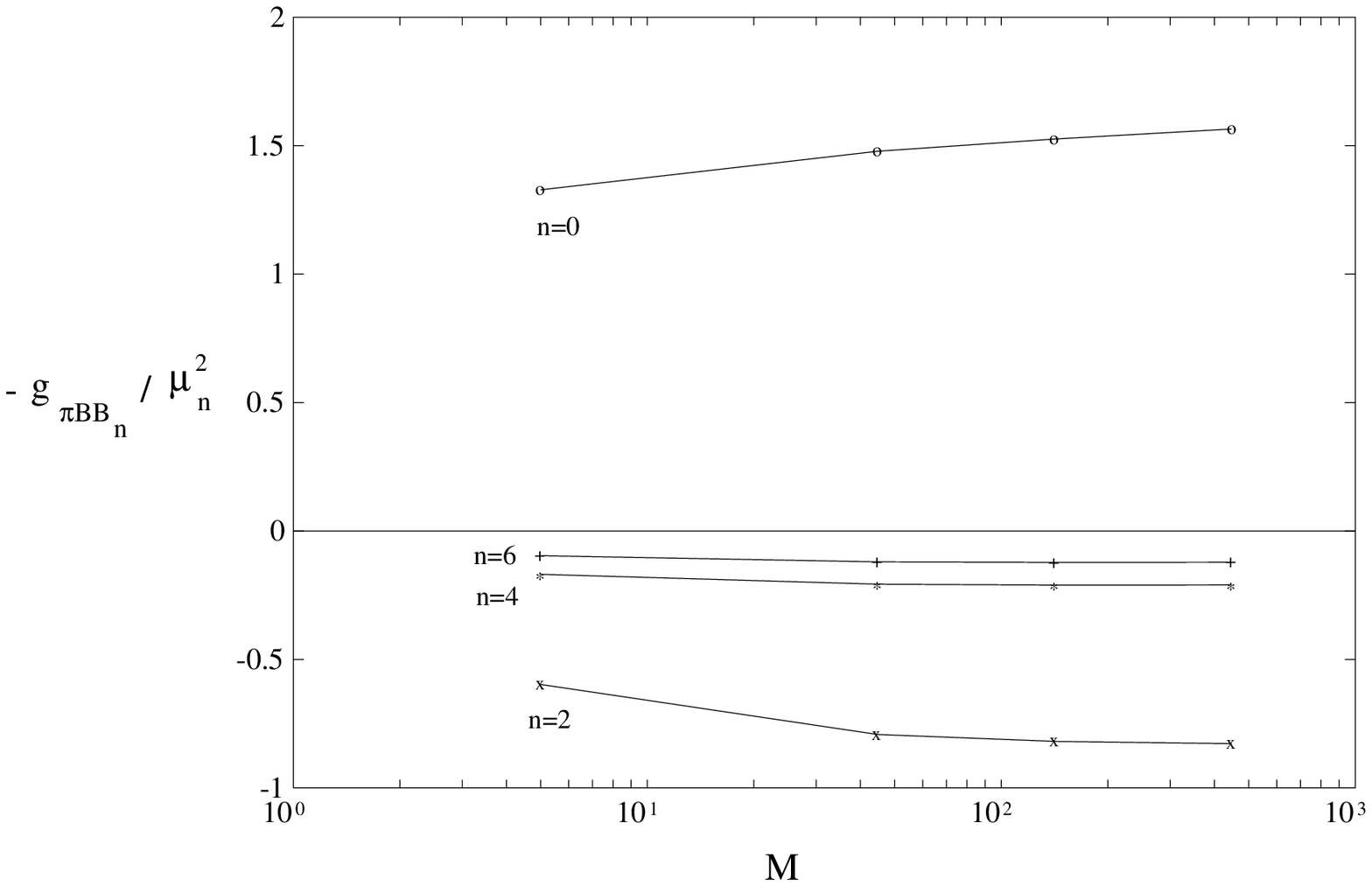}
\fcaption{Approach to the heavy quark limit:
$-\gpiBB/\mu_n^2$ vs. $M$ for $m=0.56$ ($\mu_\pi^2=3.09$)
and $M^2=25, 2000, 20000, 200000$.
Results for $m=0.1$ are similar.
}\label{fig:gvsM}
\end{figure}

That $\A_0(\mu_n^2)$ approaches the limit dictated by the
Callan-Triemann relation can be seen more clearly in
Fig.~\ref{fig:gvsM} which shows the dependence of
$-\hat\gpiBB=-\gpiBB/\mu_n^2$ on the large mass $M$ for fixed
`light' mass $m$. The dependence on the heavier mass $M$, for large
$M$, of the decay constants\cite{mendebgb} and couplings is seen to be
as expected.

\section{Conclusions}
In the 't~Hooft model the form factors of flavor changing currents for
the decay of a meson into a ``pion'' are given by a single pole in the
chiral limit. Our main result is summarized in Eq.~(\ref{finally}).

Away from the chiral limit the form factors are no longer given by a
single pole. However for quark masses $m^2<g^2N_c/\pi$ the corrections
to a single pole form factor are small, and the form factors are
effectively pole dominated over the whole physical region.

\section{Acknowledgements}
Were it not for the assistance of the organizers I would not have been
able to attend this conference. I remain greateful to them. This work
was supported in part by an Alfred P. Sloan Foundation Fellowship and
by the Department of Energy under grants DE--FG05--92ER--40722 and 
DOE--FG03--90ER40546.

%\listoffigures

\end{document}